\documentclass[a4paper]{article}

\usepackage[figuresright]{rotating}
\usepackage{graphicx}

\newcommand{\nucl}[3]{Nucl. Phys. \textbf{#1} ({#2}) {#3}.}
\newcommand{\rmp}[3]{Rev. Mod. Phys. \textbf{#1} ({#2}) {#3}.}
\newcommand{\prd}[3]{Phys. Rev. \textbf{D{#1}} ({#2}) {#3}.}
\newcommand{\plb}[3]{Phys. Lett. \textbf{B{#1}} ({#2}) {#3}.}
\renewcommand{\Box}{\partial^2}

\begin{document}

\title{Classical Solutions of SU(3) Pure Yang-Mills Theory}
\author{O. Oliveira\thanks{\textit{email}: orlando@teor.fis.uc.pt},
        R. A. Coimbra\thanks{\textit{email}: rita@teor.fis.uc.pt} \\
        Centro de {F\'\i}sica Computacional \\
        Departamento de F{\'\i}sica \\
        Universidade de Coimbra \\
        3004-516 Coimbra \\
        Portugal }

\maketitle

\begin{abstract}
Regular classical solutions of pure SU(3) gauge theories, in Minkowsky 
spacetime, are computed in the Landau gauge. 
The classical fields have an intrinsic energy scale and produce quark
confinement if interpreted in the sense of a nonrelativistic potential. 
Moreover, the quark propagator in the background of these fields
vanishes at large positive and negative time and space separations.
\end{abstract}

\section{Introduction and Motivation}

The classical solutions of a field theory are a first step towards the 
understanding of the associated quantum theory. However, for nonlinear 
theories, finding solutions of the Euler-Lagrange equations
can be particularly difficult. For nonabelian gauge theories the quest for 
classical configurations is still an open problem.

For pure gauge theories, a number of classical solutions are known when the
gauge group is SU(2) \cite{Ac79}, but even for SU(2) we do not have the 
complete picture. The relation between SU(2) and SU(3) allows to write 
classical SU(3) fields from SU(2) solutions. For QCD, classical configurations
built in this way, in particular the instanton \cite{Ho79,Ho81,Po77,ScSh98},
are used to address hadronic phenomenology. Despite the sucess of such an
approach, none of the known classical configurations is able to explain/suggest
the confinement of quarks in hadrons. Hopefully, there should be solutions of
the classical field equations which account for quark confinement and, maybe,
the chiral symmetry breaking mechanism, two of the main open problems of 
strong interaction.

The interest on classical solutions is not reduced to possible explanations of
confinement or chiral symmetry breaking. If we have such SU(3) fields one
would like to answer questions about their relevance for the quantum theory.
In particular, one would like to know if it is possible to compute the
generating functional using a semi-classical approximation or if the classical
solutions define good approximations to quark and gluon propagators.

In this paper we report on pure gauge classical SU(3) Minkowsky spacetime
solutions. The solutions are obtained using a technique inspired on Cho's
work \cite{Ch991,Ch992,ChLe99,FaNi99} and already explored for SU(2) in 
\cite{eu}. In section 2 we discuss the writing of the SU(3) gauge fields.
In section 3, a particular Landau gauge vacuum classical solution is obtained 
explicitly. The solution is regular, except at the origin, and 
suggest quark confinement. Finally, on section 4 results and conclusions are 
discussed. The appendix contains various material used along the paper.

\section{Gauge Fields for SU(3)}

For $SU(3)$ gauge theories the lagrangian reads
\begin{equation}
 \mathcal{L} \, = \, - \frac{1}{4} \, F^a_{\mu\nu} \, F^{a \, \mu\nu}
\end{equation}
where
\begin{equation}
 F^a_{\mu\nu} \, = \, \partial_\mu A^a_\nu \, - \,
                      \partial_\nu A^a_\mu  \, - \,
                      g f_{abc} A^b_\mu A^c_\nu
 \mbox{ };
\end{equation}
$A^a_\mu$ are the gluon fields. 

Let $n^a$ be a covariant constant real scalar field in the adjoint 
representation. From the definition it follows that
\begin{equation}
 D_\mu \, n ^a \, = \, 
    \partial_\mu n^a \, + \, i g \left( F^b \right)_{ac} A^b_\mu n^c
  \, = \, 0 \, ; \label{Dn1}
\end{equation}
the generators of the adjoint representation are 
$\left( F^b \right)_{ac} \, = \, -i \, f_{bac} \,$. 
Given a gluon field it is always possible to solve the above equations for 
the scalar field $n$. In this way a map from the gluon field to $n$ is
defined. Let us assume that it is possible to invert the mapping. One can
always write the gauge field us
\begin{equation}
 A^a_\mu \, = \, n^a \hat{A}_\mu \, + \, X^a_\mu \, ,
\end{equation}
where the field $X$ is orthogonal to $n$ in the sense
\begin{equation}
 n \cdot X_\mu \, = \, 0 \, .
\end{equation}
Multiplying (\ref{Dn1}) by  $\left( F^d \right)_{ea} n^d$ and solving for the
gauge fields we get
\footnote{In deriving (\ref{eqn1}) the following relation 
\begin{displaymath}
f_{abc} \, f_{dec} \, = \, \frac{2}{3} \left( \delta_{ad} \delta_{be} \, - \,
                                              \delta_{ae} \delta_{bd} \right)
          \, + \,
        \left( d_{adc} d_{bec} \, - \, d_{bdc} d_{aec} \right)
\end{displaymath}
was used.}, after some algebra,
\begin{eqnarray}
 f_{eda} \, n^d \, \partial_\mu n^a \, - \, 
         \frac{2g}{3} \, X^e_\mu n^2 \, - \,
         g \left( d_{ebh} \, d_{dch} \, - \, d_{dbh} \, d_{ech} \right) \,
                      X^b_\mu \, n^c \, n^d \, = \, 0 \, ,
 \label{eqn1}
\end{eqnarray}
where
\begin{equation}
 d_{abc} \, = \,
   \frac{1}{4}
   \mbox{Tr} \left( \lambda^a \, \left\{ \,\lambda^b \, , \,
                                           \lambda^c \, \right\}
             \right)
\end{equation}
and $\lambda^a$ are the Gell-Mann matrices. Equation (\ref{eqn1}) suggest
the following form for $X^a_\mu$,
\begin{equation}
X^a_\mu \, = \, \frac{3}{2g} \, f_{abc} \frac{n^b \, \partial_\mu n^c}{n^2}
 \, + \, Y^a_\mu
\end{equation}
with $Y^a_\mu$ verifying the constraint
\begin{equation}
   n \cdot Y_\mu \, = \, 0 \, .
\end{equation}

In terms of $\hat{A}_\mu$, $n^a$ and $Y^a_\mu$, gauge fields are given by
\begin{equation}
 A^a_\mu \, = \, \hat{A}_\mu n^a \, + \,
                 \frac{3}{2g} \, f_{abc} \,
                              \frac{n^b \, \partial_\mu n^c}{n \cdot n} \, + \,
                 Y^a_\mu
 \, ; \label{A}
\end{equation}
with $n$ and $Y$ verifying the constraints
\begin{eqnarray}
 & & n \cdot Y_\mu \, = \, 0 \, ,  \label{nY} \\
 & & D_\mu n^a \, = \, 0  . \label{Dn}
\end{eqnarray}
From (\ref{Dn}) it follows that
\begin{equation}
    n \partial_\mu n \, = \, \frac{1}{2} \, \partial_\mu n^2 \, = \, 0,
   \label{t1}
\end{equation}
and one can always choose $n^2 \, = \, 1$.

Let us look at the gauge transformation properties of $n$, $Y$ and $\hat{A}$. 
$n$ is covariant constant, therefore 
$ -i \, \left( F^c \right)_{ab} \, n^c \, \partial_\mu n^b \, / 
(n \cdot n)$ also belongs to the adjoint representation of the gauge group. 
Demanding that $Y$ is in the adjoint representation, then
\begin{equation}
   \hat{A}_\mu \, \longrightarrow \, \hat{A}_\mu \, + \,
                                     \frac{1}{g} \, 
                                         n \cdot \partial_\mu \omega
                                        \, .
\end{equation}
Constraints (\ref{nY}) and (\ref{Dn}) are scalars under gauge transformations
and parameterization (\ref{A}) provides a complete gauge invariant 
decomposition of the gluon field $A^a_\mu$.

If (\ref{A}), (\ref{nY}) and (\ref{Dn}) define a complete parameterization
of the gluon fields, the total number of independent fields on both sides of
(\ref{A}) should be the same. On the l.h.s., the number of degrees of freedom
is 16. On the r.h.s., $\hat{A}$ has 2 independent fields 
and $n$ has 8. The total number of independent fields in $Y_\mu$ is
7. In general, $Y$ can be writen as a linear combination of
linear independent fields $m^{(i)}$, $i = 1 \dots 7$, with
$m^{(i)}.n \, = \, 0$, in the adjoint representation  multiplied by 
gauge invariant vector fields $B^{(i)}$. Fields $m^{(i)}$ and
$n$ define a basis on the principal bundle associated to the gauge theory.
This parametrization introduces a large number of fields, not all are
independent.

To solve the classical equations of motion for QCD, we choose a
spherical like basis in the color space - see appendix for definitions.
Setting $n \, = \, \vec{e}_3$, condition (\ref{Dn}) becomes
\begin{equation}
 - \frac{1}{2} \partial_\mu n^a \, - \, g f_{abc} \, Y^b_\mu \, n^c \, = \, 0
 \, .
\label{Dn2}
\end{equation}
This set of equations provides the following relations between the $Y^a_\mu$ 
fields
\begin{eqnarray}
 & &  Y^2_\mu \, = \, - \, Y^1_\mu \, \cot \theta_2 \, , \label{Dn01}  \\
 & &  Y^3_\mu \, = \, - \, \frac{1}{2g} \, \partial_\mu \theta_2 \, ,  \\
 & &  Y^4_\mu \, = \, 0 \, ,                                           \\
 & &  Y^5_\mu \, = \, 0 \, ,                                           \\
 & &  Y^6_\mu \, = \, 0 \, , \\
 & &  Y^7_\mu \, = \, 0 \, . \label{Dn02}
\end{eqnarray}
Taking into account (\ref{Dn01}) to (\ref{Dn02}) the gluon field is given by
\begin{equation}
 \left( A^a_\mu \right) \, = \left( \begin{array}{c}
          - \sin \theta_2 \, \hat{A}_\mu \, + \, Y^1_\mu \\
            \cos \theta_2 \, \hat{A}_\mu \, - \, \cot \theta_2 \, Y^1_\mu \\
            \partial_\mu \theta_2 \, / \, g  \\
            0 \\
            0 \\
            0 \\
            0 \\
            Y^8_\mu
                                    \end{array} \right) \, .
\end{equation}
Condition (\ref{nY}) simplifies the gluon field into
\begin{equation}
 A^a_\mu \, = \, n^a \hat{A}_\mu \, + \, 
                 \delta^{a3} \, \frac{1}{g} \, \partial_\mu \theta_2
                 \, + \, \delta^{^a8} C_\mu \, .
 \label{glue}
\end{equation}
Then, the corresponding gluon field tensor components are
\begin{equation}
  F^a_{\mu \nu} \, = \, n^a \, \mathcal{A}_{\mu\nu} \, + \,
           \delta^{a8} \, \mathcal{C}_{\mu \nu}
\end{equation}
where
\begin{equation}
 \mathcal{A}_{\mu \nu} \, = \, \partial_\mu \hat{A}_\nu \, - \,
                               \partial_\nu \hat{A}_\mu \, , \hspace{0.7cm}
 \mathcal{C}_{\mu \nu} \, = \, \partial_\mu C_\nu \, - \,
                               \partial_\nu C_\mu \, .
\end{equation}
The classical action is a functional of the ``photon like'' fields
$\hat{A}_\mu$ and $C_\mu$,
\begin{equation}
  \mathcal{L} \, = \, - \frac{1}{4} \left(
    \mathcal{A}^2 \, + \, \mathcal{C}^2 \right) \, ,
\end{equation}
and the classical equations of motion
\begin{equation}
 D^\nu F^a_{\mu \nu} \, = \, 
    \partial^\nu F^a_{\mu \nu} \, - \,
    g f_{abc} A^b_\nu F^c_{\mu \nu} \, = \, 0
\end{equation}
are reduced to
\begin{eqnarray}
 \partial_\nu \mathcal{A}^{\mu \nu} \, = \, 0 \, ,
 & &
 \partial_\nu \mathcal{C}^{\mu \nu} \, = \, 0 \, .
 \label{equacoes}
\end{eqnarray}

The ansatz (\ref{glue}) maps SU(3) to a set of two linear theories making
classical pure SU(3) gauge theory formally equivalent to QED with
two ``photon fields''. Note that the  ``photon fields'' do not couple to each
other but their coupling to fermionic fields require different Gell-Mann
matrices - the group structure of (\ref{glue}) is SU(2)$\oplus$U(1), with
$\hat{A}_\mu$ related to SU(2) and $C_\mu$ to U(1). 
The solutions of the classical equations of motion (\ref{equacoes}) give, for
$\hat{A}_\mu$ and $C_\mu$ fields, QED-like configurations but provide
no information on the nature of $\theta_2$. As usual in QED, to compute
classical configurations we have to choose a particular gauge.

\section{Classical Solutions in Landau Gauge}

The choice of the Landau gauge is  motivated by covariance. Landau gauge is
defined by
\begin{equation}
  \partial^\mu A^a_\mu \, = \, 0 \, .
\end{equation}
In terms of $\hat{A}_\mu$, $C_\mu$ and $\theta_2$ the above conditions
read
\begin{eqnarray}
 \partial^\mu \hat{A}_\mu \, = \, 0, &  & 
             \partial^\mu C_\mu \, = \, 0  \, , \nonumber \\
 (\partial^\mu \theta_2) \hat{A}_\mu \, = \, 0  \, , & &
              \Box \theta_2 \, = \, 0 \, .
 \label{landau}
\end{eqnarray}
Instead of solving (\ref{landau}) together with (\ref{equacoes}), we look at
the vacuum solutions.
Taking the usual definitions for the chromoelectric $\vec{E}^a$ and
chromomagnetic $\vec{B}^a$ fields, the energy of a classical configuration
is independent of $\theta_2$,
\begin{eqnarray}
 H \, = & &
   \int d^4 x \, \left\{
  \frac{1}{2} \left( \vec{B}^a \cdot \vec{B}^a \, - \,
                     \vec{E}^a \cdot \vec{E}^a \right) \, - \,
  \vec{E}^a \cdot \frac{\partial \vec{A}^a}{\partial t}
  \right\} \nonumber \\
   = & &
   \int d^4 x \, \, 
  \frac{1}{2} \left( \vec{B}^a \cdot \vec{B}^a \, + \,
                     \vec{E}^a \cdot \vec{E}^a \right) \, .
\end{eqnarray}
Vacuum solutions have zero energy and
the configurations defined by $\hat{A}_\mu \, = \, C_\mu \, = \, 0$, i.e. 
$g \, A^a_\mu \, = \, \delta^{a3} \partial_\mu \theta_2$, belong to this class.
Then, (\ref{landau}) reduces to
\begin{equation}
   \Box \theta_2 \, = \, 0 \, ;
 \label{teta}
\end{equation}
$\theta_2$ does not verify any special kind of boundary condition.
The above equation can be solved by separation of variables in the usual way.
In spherical coordinates, the solution is
\begin{eqnarray}
    \theta_2 \, = \, \sum\limits^{+ \infty}_{l = 0} \, 
                     \sum\limits^{l}_{m = -l} \, 
      \left( \frac{ }{ } \right.
  & &
                     \int\limits^{+ \infty}_0 d \omega \,
                     \left\{ \frac{}{} \right.
 \,
  \left[ \frac{}{} a (\omega) e^{- i \omega t} \, + \,
         a^* (\omega) e^{+ i \omega t}  \right] \,  \times
  \nonumber \\
  & & \hspace{2.7cm}
     \left[ \frac{}{} A_{lm} ( \omega ) j_l ( \omega r ) \, + \,
            B_{lm} ( \omega ) n_l ( \omega r ) \right] \, \times
   \nonumber \\
  & & \hspace{3cm}
      Y_{lm} ( \theta , \phi )
     \nonumber \\
 & & \hspace{1.5cm} + \,
  \left[ \frac{}{}  b (\omega) e^{-  \omega t} \, + \,
         c (\omega) e^{+  \omega t}  \right] \,\times
  \nonumber \\
  & & \hspace{2.7cm}
     \left[ \frac{}{} C_{lm} ( \omega ) j_l ( i \omega r ) \, + \,
            D_{lm} ( \omega ) n_l ( i \omega r ) \right] \, \times
   \nonumber \\
  & & \hspace{3cm}
      Y_{lm} ( \theta , \phi )
  \, \left. \frac{}{} \right\}
 \nonumber \\
 & & + \,
  \left[ \frac{}{} e t \, + \, f \right] 
     \left[ F_{lm} r^l \, + \,
            \frac{G_{lm}}{r^{l+1}} \right]
      Y_{lm} ( \theta , \phi )
   \left. \frac{}{} \right) \, , \nonumber \\
 & &
 \label{theta}
\end{eqnarray}
where $\omega$ has dimensions of mass, $j_l$ is the spherical Bessel function
of order $l$ and $n_l$ is the spherical Neumann function of order $l$. 
It is curious that solution (\ref{theta}) is completely determined by the
gauge fixing condition. 

In $\theta_2$ and $A^a_\mu$ the first (proportional to $a( \omega )$) 
and third (proportional to $e$ and $f$) terms remind classical QED
configurations. The second term proportional to $b ( \omega )$ and
$c ( \omega )$ shows exponential behaviour both in space and time. 
From the point of view of possible mechanisms for quark
confinement, it is precisely this term which makes the solution
interesting. Defining the quark propagator $S$ in the usual way, 
\begin{eqnarray}
  S^{-1} \, = \, i \gamma^\mu D_\mu \, - \, m \, ,  \\
  D_\mu \, = \, \partial_\mu \, + \, i g A^a_\mu \frac{\lambda^a}{2}
  \, ,
\end{eqnarray}
the asymptotic behaviour of the terms proportional to $b ( \omega )$ and
$c ( \omega )$ in (\ref{theta}) make
\begin{equation}
 S \, \longrightarrow \, 0 \, ,
 \hspace{1cm} \mbox{when }
 t \rightarrow  \pm \infty, \,\,
 r \rightarrow \infty \, ,
\end{equation}
i.e. quarks can not propagate over large times and large space separations.
Moreover, one can build a nonrelativistic potential in the way described in
\cite{Yn99}. In first approximation, the potential is given by the time
component of the gluon field. Including only the monopole term, the
nonrelativistic potential is
\begin{equation}
 V (r; \omega ) \, = \, A \frac{ \sin ( \omega r ) }{r} \, + \, 
                        B \frac{ \cos ( \omega r ) }{r} \, + \,
                        C \frac{ \sinh ( \omega r ) }{r} \, + \,
                        D \frac{ \cosh ( \omega r ) }{r} \, .
 \label{vr}
\end{equation}
This a confining potential which for short distances reproduces a Coulomb like
plus linear potential
\begin{equation}
 V(r) \, = \, \omega \left\{
          \frac{B+D}{\omega r} \, + \, \left( A + C \right) \, + \,
          \frac{D-B}{2} \left( \omega r \right) \, + \,
          \mathcal{O} \left( \left( \omega r \right) ^2 \right)
          \right\}  \, .
 \label{potenciallinear}
\end{equation}
The various parameters in $V(r; \omega)$ can be estimated comparing the 
potential with the Cornell potential \cite{Cornell}. Assuming that in the
range $0.2 - 0.7$ fm one can use the form (\ref{potenciallinear}) and 
considering the quantity $R_0$ defined in \cite{R0}, then
$\omega \, = \, 383$ MeV, $B \, = \, -1.505$ and $D \, = \, 0.985$. 
A way to fix $A$ and $C$ is to minimise the integral of the difference squared
between (\ref{vr}) and Cornell potential for $r$ between
$0.2$ fm and $0.7$ fm; then $A \, = \, -0.069$ and $C \, = \, 0.065$. 
The two potentials are ploted in figure 1. The agreement up to distances of
$\sim 1.6$ fm is perfect. Therefore, for systems with dimension up to
$1.6$ fm, such us heavy quark systems, both potentials reproduce the same
physics. It is only when the dimension of the quark system is close or larger
than $1.6$ fm that one start to distinguish between the potentials. 
Unfortunately, the only hadron which fits such requirements is the pion and
the idea of describing the pion via a nonrelativistic potential is
questionable. 

\begin{figure}[t]
\scalebox{0.5}[0.3]{\includegraphics{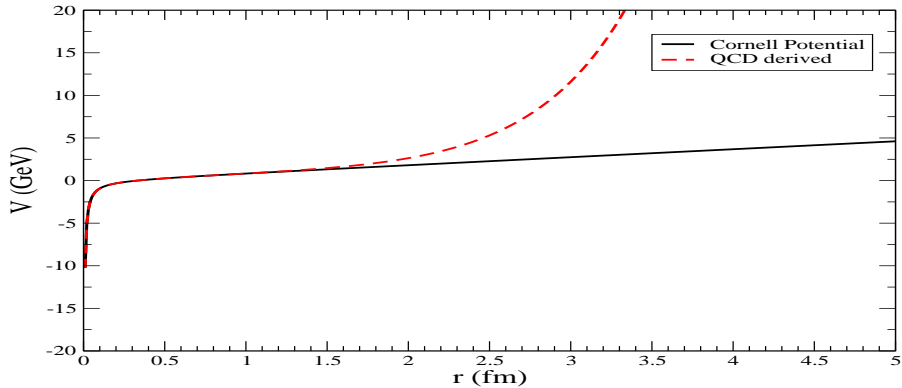}}
\caption{The Cornell potential and the potential (\ref{vr}) with
$A \, = \, -0.069$, $B \, = \, -1.505$, $C \, = \, 0.065$, $D \, = \, 0.985$
and $\omega \, = \, 383$ MeV.}
\end{figure}

The potential (\ref{vr}) has one parameter with mass dimension $\omega$.
It is curious that the value obtained by a comparisation with the Cornell
potential is within typicall values of $\Lambda_{\mbox{QCD}}$.

Another interesting class of solutions is provided by setting
\begin{equation}
 \hat{A}_\mu \, = \, \partial_\mu \chi \, 
 \hspace{0.5cm} \mbox{and} \hspace{0.5cm}
 C_\mu \, = \, \partial_\mu \xi \, ;
 \label{sol2}
\end{equation}
the classical action is again null. The equations of motion in the Landau gauge
are now reduced to
\begin{eqnarray}
 \Box \chi \, = \, 0, &  & 
             \Box \xi \, = \, 0  \, , \nonumber \\
 (\partial^\mu \theta_2) \partial_\mu \chi \, = \, 0  \, , & &
              \Box \theta_2 \, = \, 0 \, ,
 \label{landau2}
\end{eqnarray}
which have solutions with essentially the same caractheristics as
(\ref{theta}).

It is true that the classical gluon field solutions computed in this paper do
not prove  quark confinement, but seem to point in the right direction.
They support the idea that quarks do not propagate over large times (no free
quarks) and that they can not propagate over large space separations (quarks
are confined within regions of size $r \, \sim \, 1 / \omega$).

\section{Results and Conclusions}

In this paper we report vacuum solutions of classical pure SU(3) gauge theory
in Minkowsky space. The solutions were obtained after writing the gluon field
in a particular way (\ref{A}) and choosing a spherical like basis in color
space. The two steps seem crucial to map SU(3) to abelian like theories, i.e.
to replace a nonlinear theory by a set of linear field equations. 
We do not know if, in general, is always possible to build such mappings 
and if the gluon field is too strongly constrained by our procedure. 
Answering such questions requires further investigations.

In what concerns Particle Physics, the configurations 
(\ref{theta}) and (\ref{sol2}) are interesting. They are regular
in time and in space, except at the origin. The associated gluon field
diverges exponentially for large space or time separations. It is precisely
this last property which makes the solutions appealing from the point of view
of a possible explanation for quark confinement. However, one should remember 
that the configurations considered are solutions of the classical equations
of motion and their relevance to the quantum theory should be investigated
before drawing conclusions.

In order to explore for possible features related to confinement, a
nonrelativistic potential  describing heavy quarks interactions (\ref{vr})
was introduced and compared to the Cornell potential. The potentials 
match perfectly for distances up to $\sim 1.6$ fm. For larger distances, 
(\ref{vr}) diverges exponentially  and not linearly as in the Cornell
potential. The potential (\ref{vr}) is described by a mass parameter 
$\omega$ with a value compatible with $\Lambda^{n_f = 0}_{\mbox{QCD}}$. 
The relation of (\ref{vr}) to Cornell potential and between $\omega$ and
$\Lambda^{n_f = 0}_{\mbox{QCD}}$ should be taken with care, but 
certainly they suggest that (\ref{theta}) and/or (\ref{sol2},\ref{landau2}) 
should play a role in Strong Interaction Physics.

\section*{Appendix}

The color space has dimension eight. For an eight dimension space, the
spherical basis requires seven angles. Considering the following seven
functions $\theta_i (x)$, $i = 1 ... 7$ we define the basis as follows
\begin{eqnarray}
 \vec{e}_1 \, = \,  & & \left(\begin{array}{l}
     \sin \theta_1 \, \cos \theta_2 \, \sin \theta_3 \, \sin \theta_4 \,
                       \sin \theta_5 \, \sin \theta_6 \, \sin \theta_7  \\
     \sin \theta_1 \, \sin \theta_2 \, \sin \theta_3 \, \sin \theta_4  \,
                       \sin \theta_5 \, \sin \theta_6 \, \sin \theta_7  \\
     \cos \theta_1 \, \sin \theta_3 \, \sin \theta_4 \,
                       \sin \theta_5 \, \sin \theta_6 \, \sin \theta_7  \\
     \cos \theta_3 \, \sin \theta_4 \,
                       \sin \theta_5 \, \sin \theta_6 \, \sin \theta_7  \\
     \cos \theta_4 \,  \sin \theta_5 \, \sin \theta_6 \, \sin \theta_7  \\
     \cos \theta_5 \, \sin \theta_6 \, \sin \theta_7   \\
     \cos \theta_6 \, \sin \theta_7  \\
     \cos \theta_7
                            \end{array}  \right) \\
 \vec{e}_2 \, = \,  & & \left(\begin{array}{l}
     \cos \theta_1 \, \cos \theta_2  \\
     \cos \theta_1 \, \sin \theta_2  \\
     - \sin \theta_1                 \\
      0                              \\
      0                              \\
      0                              \\
      0                              \\
      0         
                            \end{array}  \right) \\
 \vec{e}_3 \, = \,  & & \left(\begin{array}{l}
     - \sin \theta_2                 \\
     \cos \theta_2                   \\
     0                               \\
      0                              \\
      0                              \\
      0                              \\
      0                              \\
      0         
                            \end{array}  \right) \\
 \vec{e}_4 \, = \,  & & \left(\begin{array}{l}
     \sin \theta_1 \, \cos \theta_2 \, \cos \theta_3  \\
     \sin \theta_1 \, \sin \theta_2 \, \cos \theta_3  \\
     \cos \theta_1 \, \cos \theta_3                   \\
      - \sin \theta_3                              \\
      0                              \\
      0                              \\
      0                              \\
      0         
                            \end{array}  \right) \\
 \vec{e}_5 \, = \,  & & \left(\begin{array}{l}
     \sin \theta_1 \, \cos \theta_2 \, \sin \theta_3 \, \cos \theta_4 \\
     \sin \theta_1 \, \sin \theta_2 \, \sin \theta_3 \, \cos \theta_4 \\
     \cos \theta_1 \, \sin \theta_3 \, \cos \theta_4                   \\
     \cos \theta_3 \, \cos \theta_4                              \\
      - \sin \theta_4                              \\
      0                              \\
      0                              \\
      0                             \end{array}  \right) \\
 \vec{e}_6 \, = \,  & & \left(\begin{array}{l}
     \sin \theta_1 \, \cos \theta_2 \, \sin \theta_3 \, \sin \theta_4
                   \, \cos \theta_5 \\
     \sin \theta_1 \, \sin \theta_2 \, \sin \theta_3 \, \sin \theta_4 
                   \, \cos \theta_5 \\
     \cos \theta_1 \, \sin \theta_3 \, \sin \theta_4 \, \cos \theta_5   \\
     \cos \theta_3 \, \sin \theta_4 \, \cos \theta_5                \\
     \cos \theta_4 \, \cos \theta_5                              \\
      - \sin \theta_5                              \\
      0                              \\
      0                             \end{array}  \right) \\
 \vec{e}_7 \, = \,  & & \left(\begin{array}{l}
     \sin \theta_1 \, \cos \theta_2 \, \sin \theta_3 \, \sin \theta_4
                   \, \sin \theta_5 \, \cos \theta_6 \\
     \sin \theta_1 \, \sin \theta_2 \, \sin \theta_3 \, \sin \theta_4 
                   \, \sin \theta_5 \, \cos \theta_6 \\
     \cos \theta_1 \, \sin \theta_3 \, \sin \theta_4 \, \sin \theta_5  
                   \, \cos \theta_6 \\
     \cos \theta_3 \, \sin \theta_4 \, \sin \theta_5 \, \cos \theta_6     \\
     \cos \theta_4 \, \sin \theta_5 \, \cos \theta_6            \\
     \cos \theta_5 \, \cos \theta_6                              \\
      - \sin \theta_6                              \\
      0                             \end{array}  \right) \\
 \vec{e}_8 \, = \,  & & \left(\begin{array}{l}
     \sin \theta_1 \, \cos \theta_2 \, \sin \theta_3 \, \sin \theta_4
                   \, \sin \theta_5 \, \sin \theta_6 \, \cos \theta_7 \\
     \sin \theta_1 \, \sin \theta_2 \, \sin \theta_3 \, \sin \theta_4 
                   \, \sin \theta_5 \, \sin \theta_6 \, \cos \theta_7 \\
     \cos \theta_1 \, \sin \theta_3 \, \sin \theta_4 \, \sin \theta_5  
                   \, \sin \theta_6 \, \cos \theta_7 \\
     \cos \theta_3 \, \sin \theta_4 \, \sin \theta_5 \, \sin \theta_6 
                   \, \cos \theta_7    \\
     \cos \theta_4 \, \sin \theta_5 \, \sin \theta_6 \, \cos \theta_7   \\
     \cos \theta_5 \, \sin \theta_6 \, \cos \theta_7                \\
     \cos \theta_6 \, \cos \theta_7                             \\
      - \sin \theta_7                           \end{array}  \right)
\end{eqnarray}



\begin{thebibliography}{99}

\bibitem{Ac79} A. Actor, \rmp{51}{1979}{461}

\bibitem{Ho79} G. 't Hooft, \nucl{B153}{1979}{141}

\bibitem{Ho81} G. 't Hooft, \nucl{B190[FS]}{1981}{455}

\bibitem{Po77} A. Polyakov, \nucl{B120}{1977}{429}

\bibitem{ScSh98} T. Schafer, E. V. Shuryak, \rmp{70}{1998}{323}

\bibitem{Ch991} Y. M. Cho, \prd{62}{2000}{074009}

\bibitem{Ch992} Y. M. Cho, J. Korean Phys. Soc. \textbf{38} (2001) 15 and
hep-th/9906198.

\bibitem{ChLe99} Y. M. Cho, H. Lee, D. G. Pak, hep-th/9905215.

\bibitem{FaNi99} L. Faddeev, A. J. Niemi, \plb{464}{1999}{90}

\bibitem{eu} O. Oliveira, hep-th/0105222. See also \cite{Li00}.

\bibitem{Li00} S. Li, Y. Zhang, Z. Zhu, \plb{487}{2000}{201}


\bibitem{Yn99} F. J. Yndur\'ain, \textit{The Theory of Quark and Gluon
Interactions}, Springer 1999.

\bibitem{Cornell} E. Eichten, K. Gottfried, T. Kinoshita, K. D. Lane,
T. M. Yan, Phys. Rev. \textbf{D21}(1980)203.

\bibitem{R0} R. Sommer, Nucl. Phys. \textbf{B411}(1994)839-854.

\end{thebibliography}
\end{document}